\documentclass[]{emulateapj}
\usepackage{natbib,amsmath}

\begin{document}

\def \npair  {74}
\def \nlarge  {49}
\def \nboss  {25}
\def \nmgii  {3X}
\def \zem {$z_{\rm em}$}
\def \zbg {$z_{\rm bg}$}
\def \zfg {$z_{\rm fg}$}
\def \mzfg {z_{\rm fg}}
\def \zlya {$z_{\rm Ly\alpha}$}
\def \wlya {$W_{\rm Ly\alpha}$}
\def \mwlya {W_{\rm Ly\alpha}}
\def \rphys {$R_\perp$}
\def \mrphys {R_\perp}
\def \lbol {$L_{\rm Bol}$}
\def \guv {$g_{\rm UV}$}
\def \nhi  {$N_{\rm HI}$}
\def \mnhi  {N_{\rm HI}}
\def \kms  {\, km~s$^{-1}$}
\def \mkms  {{\rm km~s^{-1}}}
\def \lya  {Ly$\alpha$}
\def \mlya  {{\rm Ly\alpha}}
\def \lyb  {Ly$\beta$}
\def \hMpc      {h^{-1}{\rm\ Mpc}}
\def \msol      {{\rm\ M}_\odot}
\def\cm#1{\, {\rm cm^{#1}}}
\def \cgsflux   {{\rm erg\ s^{-1}\ cm^{-2}}}
\def \cgssflux   {{\rm erg\ s^{-1}\ Hz^{-1} cm^{-2}}}
\def\sci#1{{\; \times \; 10^{#1}}}
\def\N#1{{N({\rm #1})}}

\title{A Substantial Mass of Cool,
  Metal-Enriched Gas Surrounding the Progenitors of Modern-Day
  Ellipticals}
% v3.4

\author{
J. Xavier Prochaska\altaffilmark{1,2,3},
Joseph F. Hennawi\altaffilmark{2,3},
Robert A. Simcoe\altaffilmark{4},
}
\altaffiltext{1}{Department of Astronomy and Astrophysics, UCO/Lick
  Observatory, University of California, 1156 High Street, Santa Cruz,
  CA 95064}
\altaffiltext{2}{Max-Planck-Institut f\"ur Astronomie, K\"onigstuhl}
\altaffiltext{3}{Visiting Astronomer, W.M. Keck Telescope.
The Keck Observatory is a joint facility of the University
of California and the California Institute of Technology.}
\altaffiltext{4}{MIT Kavli Institute for Astrophysics and Space Research}

\begin{abstract}
The hosts of luminous $z\sim 2$ quasars evolve into today's massive
elliptical galaxies. Current theories predict that the circum-galactic
medium (CGM) of these massive, dark-matter halos ($M_{\rm DM} \sim 
 10^{12.5} \msol$) should be dominated by a $T\sim 10^{7}\,K$
 virialized plasma. We test this hypothesis with observations of
 \npair\ close-projected quasar pairs, using spectra of the
 background QSO to characterize the CGM of the foreground
 one. Surprisingly,  our measurements reveal a cool ($T\approx
 10^4$\,K), massive ($M_{\rm CGM}> 10^{10} \msol$), and 
 metal-enriched ($Z\gtrsim 0.1 Z_\odot$) medium extending to at least
 the expected virial radius ($r_{\rm vir} = 160$\,kpc).  The average
 equivalent widths of \ion{H}{1} \lya\ ($\bar W_{\rm 
   Ly\alpha} = 2.1 \pm 0.15$\AA\ for impact parameters $\mrphys<
 200$\,kpc) and \ion{C}{2}~1334 ($\bar W_{1334} = 0.7 \pm 0.1$) exceed
 the corresponding CGM measurements of these transitions from 
 all galaxy populations studied previously.  Furthermore, we
 conservatively estimate that the quasar CGM has a $64^{+6}_{-7}\%$
 covering fraction of optically thick gas ($N_{\rm HI}>
 10^{17.2}\,{\rm cm}^{-2}$) within $r_{\rm vir}$; this covering factor
 is twice that of the contemporaneous Lyman Break Galaxy
 population.  This unexpected reservoir of cool gas is rarely detected
 ``down-the-barrel'' to quasars, and hence it is likely that our
 background sightlines intercept gas which is shadowed from the quasar
 ionizing radiation by the same obscuring medium often invoked in
 models of AGN unification.  Because the high-$z$ halos 
 inhabited by quasars predate modern groups and clusters, these
 observations are also relevant to the formation and enrichment
 history of the intragroup/intracluster medium.
\end{abstract}


\section{Introduction}

Quasars' large-scale clustering properties imply that active nuclei
reside on average in dark matter halos with $M \approx 10^{12.5}
\msol$ at $z \approx 2$ \citep[e.g.][]{pn06,white12} making them
signposts for massive galaxies at high-redshift. Therefore, their
associated host galaxies should preferentially evolve into massive and
luminous red elliptical galaxies at $z=0$ \citep{white12}.
%% JFH Does Charlie have a paper on this. It is discussed in White et al.
%% so you should also cite that paper. 
%% JXP He does not, I fear.
%% JFH Okay I added White as the reference. 
Today, such objects are typically found in groups or clusters
containing tens to thousands of galaxies.  These systems are embedded
within a hot ($T> 10^7$\,K),
tenuous plasma of virialized gas, termed the intragroup or intracluster
medium (IGrM/ICM), which dominates the halo's baryonic mass
\citep[e.g.][]{ars+08,dai10}.  At $z>1$, the IGrM/ICM becomes observationally
difficult to characterize or even detect:  X-ray telescopes are
challenged by declining surface brightness with redshift, 
Sunyaev-Zel’dovich surveys are just beginning to locate 
objects at $z>0.5$ \citep[e.g.][]{reichardt12,song12}, and neither of
these methods is sensitive to nascent clusters whose ICM has not yet
shocked into a high-temperature state.

We have recently exploited absorption spectroscopy of {\em background}
(b/g) quasars to study the diffuse gas surrounding randomly intercepted {\em
  foreground} (f/g) quasars, and by extension the massive galaxies that host
them.  In the context of individual field galaxies, this gas is now
routinely referred to as a circum-galactic medium (CGM), but if
massive quasar hosts trace group/cluster environments, it must also be
closely related to the evolving properties of the IGrM/ICM.

\begin{figure*}
\begin{center}
\includegraphics[angle=90,width=6.7in,bb=116 7 599 775]{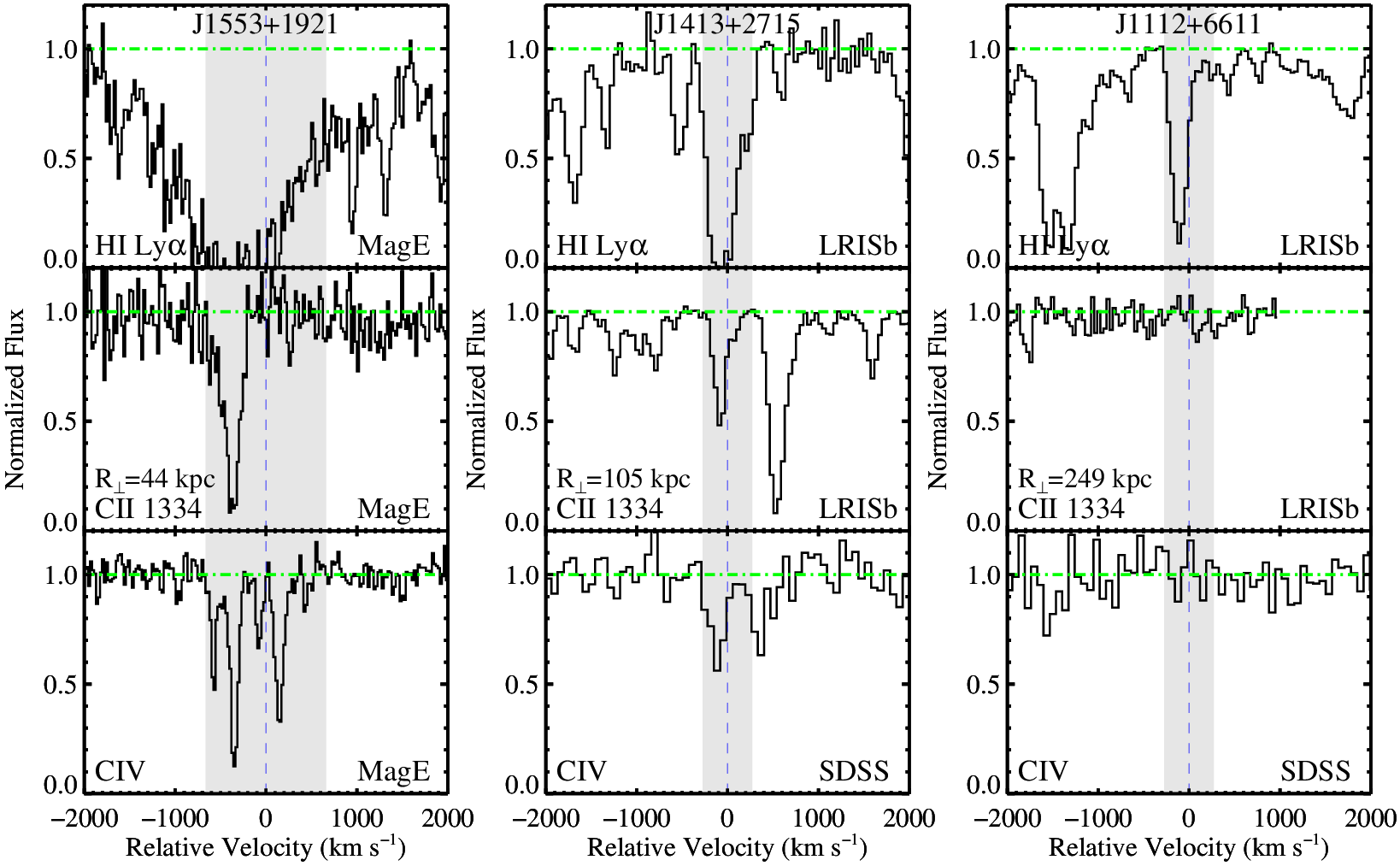}
\caption{Velocity profiles of the \ion{H}{1} \lya, \ion{C}{2}~1334,
  and \ion{C}{4} 1548/1550 transitions displayed relative to \zfg, for three
  representative quasar
  pairs.  The shaded region shows the $1\sigma$ error
  estimate for \zfg.  All of these examples exhibit strong \lya\
  absorption ($\mwlya> 1$\AA) at $z\approx \mzfg$, including a DLA at
  $\mrphys= 44$\,kpc from J1553+1921.  The first two pairs also
  exhibit strong \ion{C}{2}~1334 absorption consistent with the \lya\
  line and modest \ion{C}{4} absorption.%, whereas J1112+6611 at $\mrphys
%  = 249$\,kpc exhibits no significant metal-line absorption.  
}
\label{fig:ex_spec}
\end{center}
\end{figure*}

Absorption-line CGM measurements of nearby field galaxies universally
detect strong
\lya\ \citep[e.g.][]{lbt+95,wakker09,pwc+11,thom12}, with
line-widths indicating a cool gas ($T< 10^5$\,K).  This material
also exhibits small velocity offset ($\sim 100\mkms$) from the
galaxies' systemic redshifts suggesting that it is gravitationally
bound.  Absorption studies of the local IGrM/ICM are few, yet suggest
that the cool CGM seen in the field is suppressed
\citep{lopez08,wakker09,yoon12}. 
This may indicate that the hot, virialized IGrM/ICM prohibits the
formation of a long-lived, cooler phase \citep[e.g.][]{mb04}.
%% JFH I think you need to be a bit more explicit about what you mean 
%% when you say the cold phase is suppressed. I can see a few possibilities: 
%% 1. the bulk of the baryons are hot T ~ 10^7 K. This would result in 
%% in negligible low-ion absorption or HI absorption, because the gas is 
%% collisionally ionized. I don't know what thermal ionization is?
%% 2. A significant enough fraction of the gas is hot such that transport
%% processes, such as conduction, or turbulent mixing due to hydrodynamic instabilities
%% either destroy cold clouds T ~ 10^4 K clouds or prevent their formation in the
%% first place. You need to say these things somewhere in the paper, since they are
%% crucial to your ICM/IGRM link. I mentioned them also below when you discuss quasars
%% (transport processes) but it may be more appropriate to discuss it here 
%% where you talk about clusters. 
%% [CHAT]

CGM observations of $z\sim 2$, star-forming galaxies show
qualitatively similar patterns to present-day $L^*$ field galaxies
including enhanced \ion{H}{1} absorption to $\approx 300$\,kpc and
metal-line absorption to at least 100\,kpc
\citep[LBGs;][]{steidel+10,rakic12,rudie12,crighton12}.  The detection
of metals, in particular, has motivated discussion of feedback processes
(e.g.\ supernovae-driven winds) as agents for enriching matter in the
CGM and beyond \citep[e.g.][]{steidel+10,shen+12}.  In principle such
feedback could affect the ICM in its formative period.

The quasar host halos selected by our survey are on average several to
ten times more massive than LBGs \citep{white12}, so it is conceivable
that their CGM is correspondingly more massive and would exhibit
distinct physical characteristics.  For example, if the QSO-CGM is
shock-heated to high temperatures during virialization it may lack the
cool phase traced by \ion{H}{1} absorption, either
because the majority of gas is hot, or because astrophysical transport
processes 
(e.g.\ conduction, turbulent mixing) suppress the cold phase. 
It is also possible
that the quasar may photoionize gas from tens to several hundreds of kpc
and photo-evaporate cool clumps \citep[e.g.][]{QPQ2,cmb+08}.
Finally, kinetic feedback from the AGN may stir, heat, and enrich
the CGM.

%The properties of the quasar CGM offer 
%insight into massive galaxy formation and AGN feedback at $z \approx
%2$, and also has implications for the origin and evolution of
%the IGrM/ICM.  
%Previously,  we studied a handful of small-impact
%parameter quasar pairs and 
%found a high-incidence of optically thick gas in the quasar hosts'
%CGM \citep{QPQ1}, and argued that quasars emit their ionizing
%radiation anisotropically \citep{QPQ2}.  
%Here we present results on the average absorption
%properties of $z \sim 2$ quasar host haloes using a much larger
%sample of 74 sightlines.  
%These provide an unprecedented resolution of the properties of gas
%surrounding their massive halos.
%We adopt a standard $\Lambda$CDM
%cosmology \citep{wmap05} and all distances are given in proper units.

The properties of the quasar CGM offer insight into massive galaxy
formation and AGN feedback at $z\sim 2$, and also has implications for the
origin and evolution of the IGrM/ICM.  In our Quasars Probing Quasars
(QPQ) program, absorption-line observations of
samples of projected quasars pairs are used to characterize the
quasar CGM.  We previously studied a sample of small-impact parameter
quasar pairs and found a high-incidence of optically thick gas in the
quasar hosts’ CGM \citep{QPQ1}, and argued that
quasars emit their ionizing radiation anisotropically \citep{QPQ2}. 
Detailed absorption line modeling of an echelle
spectrum of a projected quasar pair was presented in \citep{QPQ3}. 
Here we present results on the average absorption
properties of $z\sim 2$ quasar host haloes 
using a much enlarged sample of
74 sightlines. These provide an unprecedented resolution of the
properties of gas surrounding their massive halos. We adopt a standard
$\Lambda$CDM cosmology 
\citep[$\Omega_M=0.26,\Omega_\Lambda=0.74,H_0=70\mkms\,{\rm Mpc^{-1}}$;][]{wmap05} and all distances are given in proper units.

%%%%%%%%%%%%%%%%%%%%%%%%%%%%%
\section{Data}

Using data-mining techniques suited to large surveys such as the Sloan
Digital Sky Survey (SDSS),
we have identified $\approx 300$ projected pairs of quasars, with
redshift difference $\delta v> 2000\mkms$ and impact parameter
$\mrphys< 300$\,kpc \citep[e.g.][]{hso+06,QPQ1,hennawi10}.  
We obtained deep,
high-resolution spectroscopy of the b/g quasar for
\nlarge\ pairs using the Keck, Gemini, or Magellan telescopes.  For an
additional \nboss\ systems, public datasets from the SDSS and BOSS
surveys \citep{sdssdr7,boss_dr9} provide b/g spectra with
signal-to-noise $S/N$ exceeding 10 (per rest-frame \AA) at \lya, and
we use these survey data directly.

The observations and data reduction followed standard procedures;
their details are described elsewhere (Prochaska et al., in prep.).
Table~\ref{tab:qpq5_summ} lists the quasar pair sample and
summarizes several key properties of the pairs and spectral
dataset. Redshift estimates and errors for the f/g quasars were
derived as described in \cite[][see also Shen et al.\ 2007]{QPQ1}, using one or more
well-detected rest-frame UV lines. 
% such as \ion{Mg}{2}, [\ion{C}{3}],
%\ion{Si}{4}, or \ion{C}{4} with appropriate statistical corrections.
%We estimate an uncertainty of 272\,\kms\ for the \nmgii\ pairs
%where \ion{Mg}{2} emission was measured.  The remainder of the sample
%have uncertainties ranging from $500-800\mkms$.
Each b/g quasar spectrum was continuum normalized using custom
software; 
%that follows the undulations and emission-lines present in
%the quasar SED.  
we estimate a $15\%$ (5\%) uncertainty for the
normalization within (outside) the \lya\ forest.
Figure~\ref{fig:ex_spec} shows representative
velocity plots for CGM absorption associated with the f/g quasar.

\begin{figure}
\begin{center}
\includegraphics[width=3.5in, bb=37 180 377 743]{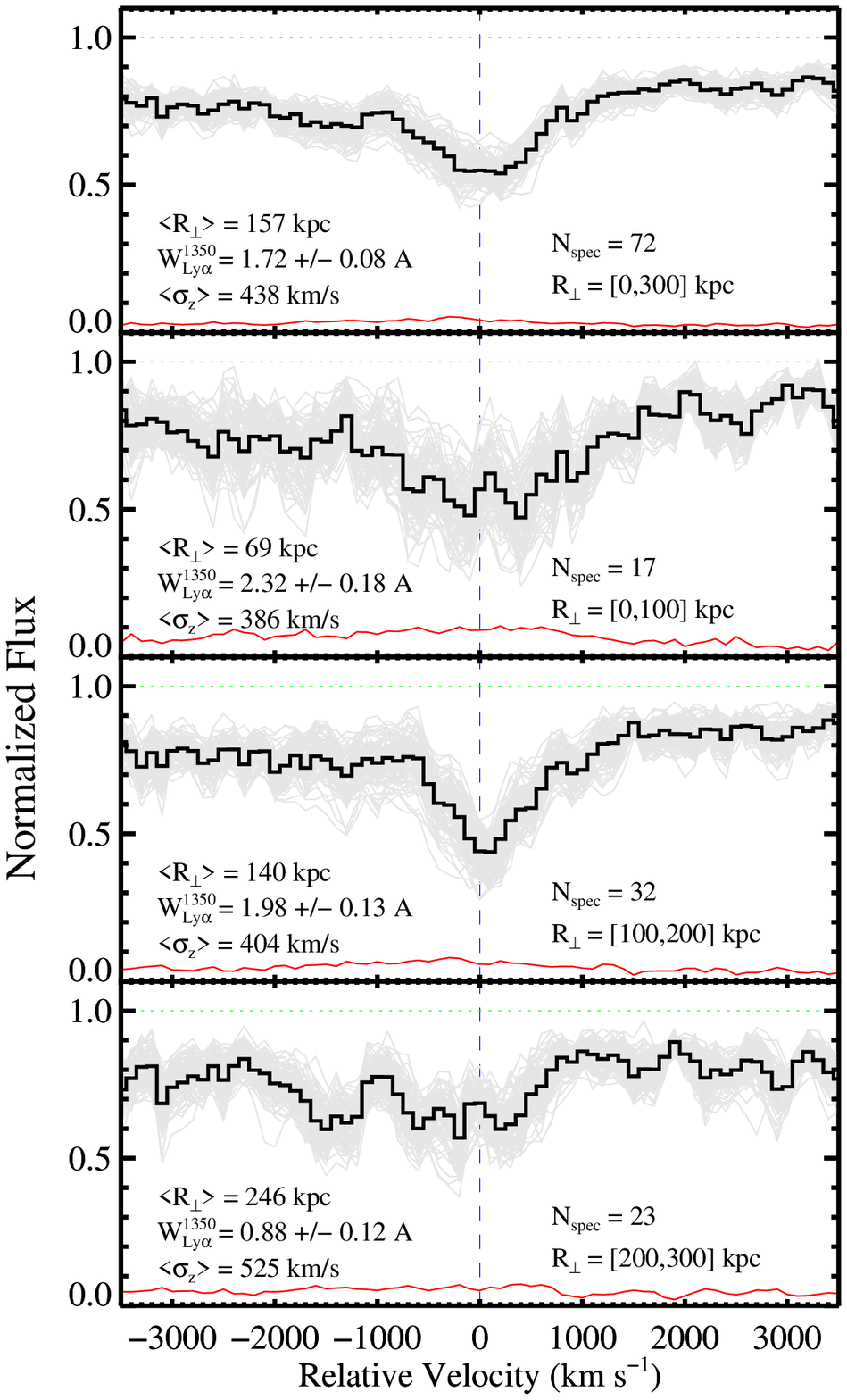}
\caption{Average \ion{H}{1} \lya\ profiles (black histogram) from
  spectra stacked at \zfg, binned by intervals of impact parameter, with
  $<\mrphys>$ the average of each set of $N_{\rm spec}$ spectra.  The
  gray lines display results for 100 random bootstrap
  realizations of the stacks and the red
  lines trace the RMS at each 100\,\kms\ pixel.  One observes 
  significant absorption at all velocities owing to the `background'
  IGM \citep[e.g.][]{kts+05}.  In each stack, one measures enhanced
  \lya\ absorption at the f/g quasar with an average equivalent width
  (integrated across $\pm 1350\,\mkms$, to encompasee the $\approx \pm
  3\sigma$ redshift uncertainty while minimizing blending with
  \ion{Si}{3}~1206, relative to the IGM-absorbed
  ``continuum'') ranging from $\mwlya^{1350}\approx 2.3$\AA\ at
  $\mrphys< 100$\,kpc to $\lesssim 1$\AA\ at $\mrphys \approx
  250$\,kpc.  In each panel, $<\sigma_z>$ reports the average quasar redshift
  uncertainty.  %The data may show broader \lya\
%  absorption at $\mrphys < 100$\,kpc, but a larger sample with smaller
%  $\sigma_z$ is required to confirm.
Note that the excess absorption at $v< -1500\mkms$ is dominated by
metal-line opacity from \ion{Si}{3}~1206.
}
\label{fig:stack}
\end{center}
\end{figure}

%%%%%%%%%%%%%%%%%%%%%%%%%%%%%
\section{Mapping the CGM of Galaxies Hosting 
$z\sim 2$ Quasars}

Our analysis focuses on \lya\ and far-UV absorption lines, which
should ideally be measured at wavelengths corresponding precisely to
each f/g quasar's redshift.  However, quasar redshifts measured from
UV emission lines have systematic uncertainties of
several hundred to 1,000\,\kms\ \citep{richards02}. 
More importantly, the $z\sim 2$ IGM exhibits a nearly continuous
\lya\ opacity which complicates attempts to assign particular
absorption lines to the CGM of a given quasar host \citep[common
practice in the low-density, $z\sim 0$ \lya\ forest;
e.g.][]{pwc+11}. 

Faced with these challenges (and motivated by analogous studies of the
field-galaxy CGM at these redshifts), we have chosen to average
(i.e.\ stack) spectra in the rest-frame of the f/g quasar for our
initial analysis.  Stacking allows us to combine data with a diversity
of spectral resolution and ${\rm S/N}$ ratio; it also averages down
the highly stochastic, background IGM
absorption to a uniform signal. Redshift errors smear the signal but
preserve its equivalent width.  The disadvantages of stacking are that
one only recovers the average equivalent width in crude bins of impact
parameter, the absorption signal could be dominated by a handful of
very strong systems (i.e.\ damped \lya\ systems, DLAs), and it is difficult
to assess gas kinematics (internal or relative to the host galaxy).

Figure~\ref{fig:stack} shows the mean continuum-normalized b/g quasar
spectrum, shifted to the rest-frame of the f/g quasar and linearly
interpolated onto a fixed velocity grid of dispersion $\Delta v =
100\mkms$.  The top panel shows the full sample, followed by bins
of stacks with increasing impact parameter.  There is significant
\ion{H}{1} \lya\ absorption relative to the IGM background, roughly
centered at \zfg\ at all impact parameters.  The average equivalent
width ranges from $\mwlya\approx 2.3$\AA\ at $\mrphys< 100$\,kpc to
$\mwlya\approx 1$\AA\ at $\mrphys\approx 300$\,kpc.  Despite their
proximity to an ultra-luminous source of ionizing radiation, quasars'
CGM exhibit strong \ion{H}{1} absorption at transverse distances
comparable to the virial radius for a $10^{12.5}M_\odot$ dark matter
halo ($r_{\rm vir}\simeq 160\,$kpc).  We have generated similar stacks
using median statistics and also with known DLA sightlines removed and
the results are qualitatively similar.  The extended haloes of quasar
hosts show strong \ion{H}{1} absorption with $\mwlya\approx 2$\AA\ to
$\mrphys= 200$\,kpc and $\mwlya\approx 1$\AA\ to $\mrphys=300$\,kpc.

These large equivalent widths suggest a correspondingly large
average \ion{H}{1} column density $\mnhi\gtrsim 10^{17} \cm{-2}$,
which would be optically thick at the Lyman limit.  For example, gas
with $\mnhi= 10^{18.7} \cm{-2}$ and a Doppler width $b=35\mkms$ has
$\mwlya= 1.7$\AA.  
Our data have sufficient ${\rm S/N}$ ratio to
characterize these individual absorbers and gain additional insight
into properties of the CGM. 
We identified the strongest absorption system in
the $\pm 1500 \, \mkms$ velocity window around \zfg\ and measured its
Ly$\alpha$ equivalent width and 
fit a Voigt profile to estimate \nhi\ (Table~\ref{tab:qpq5_summ}).  We also
searched for metal-lines in the b/g quasar's spectrum in same redshift
window, using clean spectral regions redward of the Ly$\alpha$ forest.
Objects were classified into three categories: {\it optically thick,
  ambiguous, or optically thin}, with the former showing obvious damping
wings, Lyman limit absorption, strong low-ion metal absorption ($W>
0.3$\,\AA\ 
for \ion{C}{2}~1334, \ion{O}{1}~1302,
etc.) and/or $\mwlya\ge 1.7$\,\AA.  Systems with $\mwlya\ll 1$\,\AA\ are
classified as optically thin and the remainder (a significant
population) are designated ambiguous.  
The large fraction of ambiguous cases, which may be optically thick,
means that the covering factor
$f_{\rm C}$ deduced from these data should be considered a
conservative lower limit (see Table~\ref{tab:qpq5_summ}).

%% JFH Change the virial radius to 160 kpc, and add an arrow. 
%% [CHAT]
\begin{figure}
\begin{center}
\includegraphics[width=3.5in, bb=36 167 290 656]{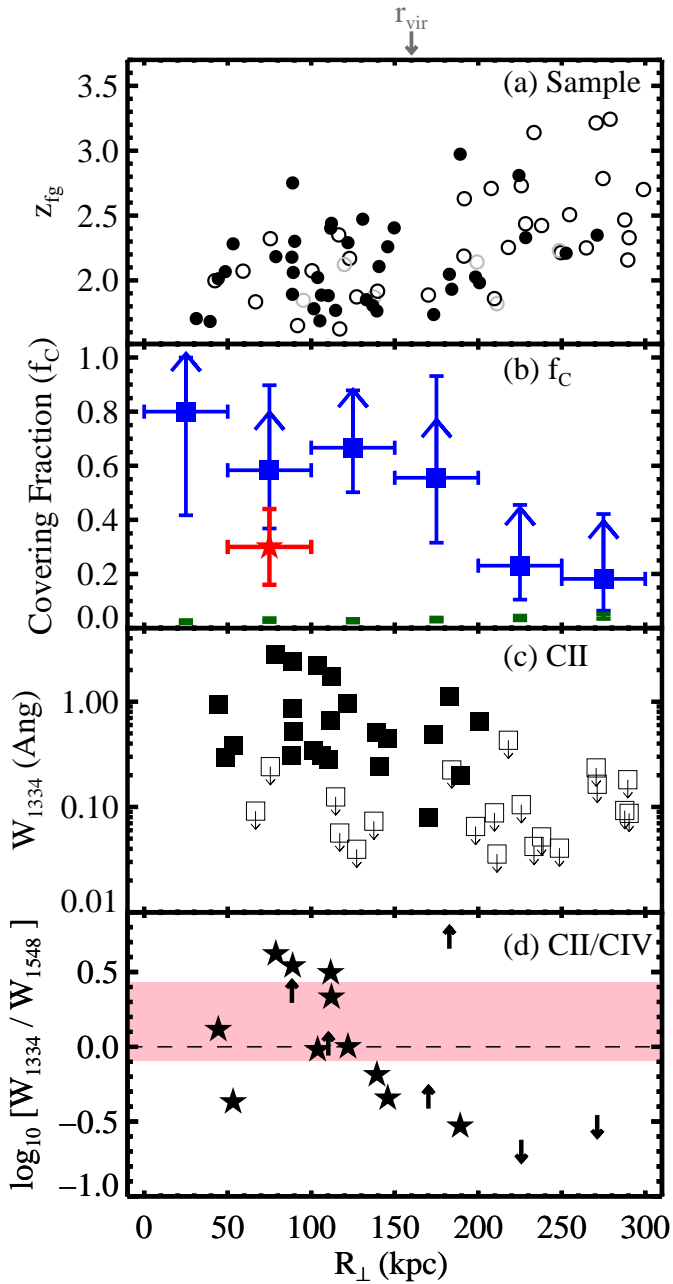}
\caption{Panel (a) shows the distribution of
  \rphys\ and \zfg\ for the sample, with filled/open/gray circles
  corresponding to optically-thick/ambiguous/optically-thin systems
  respectively;  
  (b) plots the covering fraction $f_C$ of optically
  thick gas from the CGM surrounding $z\sim 2$ quasar host galaxies in
  bins of impact parameter \rphys\ (blue points).  The covering factor is
  remarkably high ($f_{\rm C} \gtrsim 0.6$) to $\mrphys= 200$\,kpc 
  (comparable to $r_{\rm vir} \approx 160$\,kpc) and
  then drops appreciably. We also show estimates
  for $f_{\rm C}$ in $\pm 1500\mkms$ spectral regions due to the background IGM
  \citep[green points;][]{omeara+12} and an estimate for LBGs
  \citep[red point;][]{rudie12};
  (c) plots the equivalent widths measured for the \ion{C}{2}~1334
  transition.  Open symbols are non-detections, plotted at the
  $2\sigma$ value.  The preponderance of large $W_{1334}$ values  
  at $\mrphys< 200$\,kpc follows from
  the high incidence of optically thick systems and indicates an
  enriched CGM; and   
  panel (d) plots the logarithmic ratio of \ion{C}{2}~1334 and
  \ion{C}{4}~1548 equivalent widths.  At small \rphys, the CGM of
  quasar host galaxies is dominated by low-ion absorption with values
  comparable to those measured for $z>2$ DLAs \citep[pink
  region;][]{pwh+07}.  }
\label{fig:multi}
\end{center}
\end{figure}

The distribution of f/g quasar redshifts and transverse separations is
shown in the scatter plot of Figure~\ref{fig:multi}a. Filled symbols
represent quasar hosts with optically-thick absorption.  The fraction
of absorbers in this class is very high for low impact parameters; 32
out of 50 sightlines with $\mrphys < 200$\,kpc are optically thick,
corresponding to a covering factor $f_{\rm C} = 0.64^{+0.06}_{-0.07}$.  
At $\mrphys > 200$\,kpc, none of the systems is definitively optically thick (most are
ambiguous).

According to Figure~\ref{fig:multi}a, if the QSO-CGM is significantly
enriched one should also observe strong absorption from neutral or
singly ionized metal species.  Figure~\ref{fig:multi}b presents measurements
of \ion{C}{2}~1334 at the velocity of the individual strong \ion{H}{1}
lines identified as above (e.g.\ Figure~\ref{fig:ex_spec}).  As with
\ion{H}{1}, we find a preponderance of strong \ion{C}{2}~1334
absorption at $\mrphys< 200$\,kpc followed by a marked decline at
$\mrphys> 200$\,kpc.  These large equivalent widths ($W_{1334}>
0.5$\AA) must result from a combination of significant column density
and complex gas kinematics \citep[e.g.][]{QPQ3}.  Furthermore, the sharp drop in positive
detections and the coincident decline in covering fraction of
optically thick gas at $\mrphys\approx 200$\,kpc is indicative of
an association with the host galaxy, and suggests that the sampled
impact parameters circumscribe the CGM boundary of massive galaxies.

Figure~\ref{fig:multi}c shows the ratio of $W_{1334}$ to the
equivalent width of \ion{C}{4}~1548 ($W_{1548}$) for those systems where
both lines were analyzed and at least one of the two transitions was
detected.  At $\mrphys< 125$\,kpc, systems tend to have relatively
stronger \ion{C}{2} with ratios resembling those of the predominantly
neutral DLAs \citep{pwh+07}.  This suggests a medium
dominated by lower-ionization state gas, consistent with the
properties of an optically thick system.%, and it may preclude a
%predominantly warm CGM ($T \sim 10^5$\,K; e.g.\ QPQ3). 

%%%%%%%%%%%%%%%%%%%%%%%
\section{Discussion}

Using a new sample of projected quasar pair sightlines with $5\times$
more objects, we confirm strong \ion{H}{1} absorption in the
circum-galactic environment of $z\sim 2$ quasar hosts, to at least
$\mrphys= 300$\,kpc where our sample is bounded.  We further detect
strong \ion{C}{2}~1334 absorption in pairs to $\mrphys\approx 200$\,kpc,
indicating a high covering fraction ($f_C\gtrsim 0.6$) of optically
thick gas inside this radius.  These results further support our
earlier interpretation that prominent absorbing structures in the
quasars' CGM are not illuminated by the central engine \citep{QPQ2}.  We
favor scenarios where the
ionizing emission is anisotropic as predicted in AGN unification
models, as opposed to a model where the gas has not yet been
illuminated owing to the finite light-travel time required.
The 60\% covering factor exceeds similar estimates for the CGM of LBGs
\citep{rudie12} and current models for ``cold streams'' seen in
numerical
simulations of galaxy formation \citep[e.g.][]{ck11,fumagalli11a}.  

The observations require that even massive galaxies harbor a partially
cool ($T\sim 10^4$\,K) CGM, whose mass can be significant.  Assuming
a conservatively-low total gas column $N_{\rm H}$  (we assume
$10^{19}\cm{-2}$ based on our \nhi\ measurements and a modest but
highly uncertain ionization correction), with a
covering factor $f_C$ over a
projected area $\pi R^2$, we estimate

\begin{align*}
M_{\rm cool}^{\rm CGM} = \mu m_p f_c \pi R^2 N_{\rm H} 
&\approx 10^{10} (f_C/0.6) \pi (\mrphys/200\,{\rm kpc})^2 \\
&(N_{\rm H}/10^{19}\cm{-2}) \msol \;\;\; .  
\end{align*}
A more typical $N_{\rm H}$ value may exceed $10^{20}\cm{-2}$
\citep[e.g.][]{QPQ3}, implying $M_{\rm cool}^{\rm CGM}> 10^{11}\msol$.  For
a dark matter halo with $M= 10^{12.5}\msol$, the cool CGM may
easily surpass the stellar mass of the host
and could dominate the total baryons in the halo. 

One may crudely estimate gas metallicity using our \ion{C}{2}
measurements.  For $W_{1334}= 0.5$\AA, assuming 
the linear curve-of-growth yields a
very conservative lower limit of $\N{C^+}> 10^{14.5} \cm{-2}$.  For
$\mnhi= 10^{18.5} \cm{-2}$, which we believe to be typical of our
sample (Table~\ref{tab:qpq5_summ}), this implies a C/H abundance of
$\approx 1/2$ solar, ignoring ionization corrections (which lower the
estimate).  This value
matches the abundance derived for one system from a
detailed analysis using resolved metal-line absorption
\citep[J1204+0221;][]{QPQ3}.
Folding in uncertainty in $\mnhi$ and ionization, the data permit
values ranging from $Z/Z_\odot\approx 0.03 - 1.0$.  Forthcoming
analysis of our higher-resolution spectra will address the metallicity
distribution.

\begin{figure}
\begin{center}
\includegraphics[width=3.5in, bb=39 217 343 741]{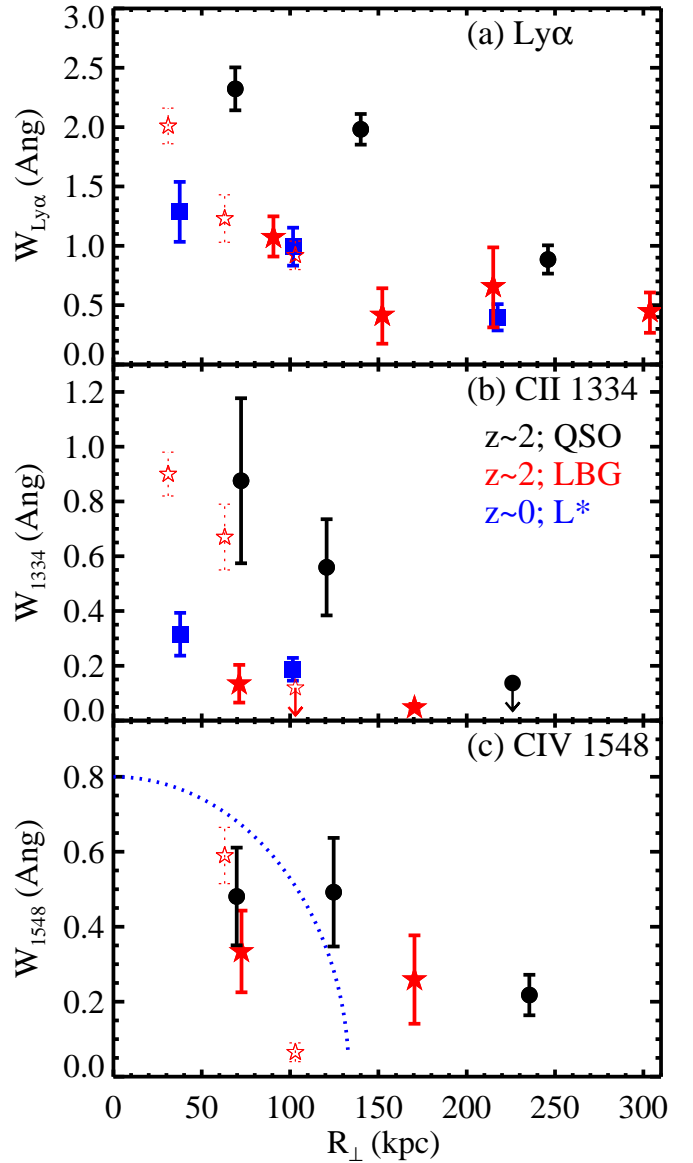}
\caption{Comparison of the average equivalent widths for (top) \ion{H}{1} \lya, (middle)
  \ion{C}{2}~1334, and (bottom) \ion{C}{4}~1548 between the results
  presented here (QSOs; black) against measurements for the $z \sim 2$
  LBGs (red) and low-$z$ $L^*$ galaxies (blue).   The CGM surrounding
  the massive galaxies hosting quasars
  exhibits much stronger \lya\ absorption at all radii,
  indicative of a more massive galaxy population.  Similarly, the
  average low-ion absorption exceeds the measurements of $L^*$
  galaxies at $z\approx 2$ and today.  
  For the LBGs, we present estimates from projected LBG/LBG pairs
  \citep[dotted][]{steidel+10} and a smaller, high-dispersion sample of projected
  quasar/LBG pairs
  \citep[solid;][]{ass+05,simcoe06,rakic12,rudie12,crighton12}.  The
  low-$z$ measurements were drawn from analyses of $L^*$
  galaxies \citep[][Tumlinson et al.\, in prep.]{clw01,pwc+11,werk12b}.
}
\label{fig:compare_cgm}
\end{center}
\end{figure}

The CGM properties of massive galaxies hosting quasars appear to be qualitatively
different from those of lower mass galaxies, in the sense that they
show stronger and more extended cool gas absorption.  
Figure~\ref{fig:compare_cgm} compares the
\ion{H}{1}, \ion{C}{2}~1334, and \ion{C}{4}~1548 statistical
absorption profiles of the sample with those of $z\sim 2.5$ LBGs.
The LBG points include both stacked spectra of projected galaxy pairs
from a large sample \citep{steidel+10}, and also detailed measurements
of individual quasar/LBG pairs \citep{simcoe06,rudie12,rakic12,crighton12},
which tend to find weaker metal-line absorption than the stacks.  The
CGM of quasar hosts exhibit significantly stronger \ion{H}{1},
\ion{C}{2}, and \ion{C}{4} absorption, especially at large radii.
Equivalent width is driven by both gas column density and kinematics,
so the larger values reflect either a more massive reservoir of cool
gas, more extreme dynamics, or both.  Whatever mechanism(s)
generate the CGM of lower mass LBGs (e.g.\ inflows, outflows), these
must be even more active in the halos surrounding quasars.
Again, these results appear to contradict the cold-flow paradigm which
predicts lower mass fractions of cool gas in more massive halos
\citep[e.g.][]{keres09a,freeke12}.  Quantitative comparisons to such predictions may require an
alternative paradigm for the gas surrounding massive galaxies.

Figure~\ref{fig:compare_cgm} also compares our results to low-$z$
$L^*$ galaxies \citep[][Tumlinson et al.\, in prep.]{clw01,werk12b}, whose CGM is
remarkably similar to that of the LBGs \citep{chen12}.  Quasar hosts,
however, are thought to evolve into massive elliptical systems resembling the
large red galaxy (LRG) population.  CGM measurements of LRGs remain
sparse, but the incidence of optically thick gas around such galaxies
\citep[traced by strong, $W_{2796}> 0.5$\AA, \ion{Mg}{2}
  absorption;][]{gauthier+10,bc11} is less than 10\%,
i.e.\ far lower than the $f_C = 0.6$ for quasar hosts.  If
LRGs are the descendants of galaxies hosting $z>2$ quasars, then their
CGM must undergo a major transformation, perhaps together with the
quenching of star-formation in the host galaxy.  

%It is tempting to relate
%this process to the formation and heating of the IGrM/ICM of clusters
%where LRGs predominantly reside, though this possibility remains
%speculative at present.
%

At $T\sim 10^4$K, the observed gas is three orders of magnitude colder than the
canonical IGrM/ICM and its entropy $S\equiv kT/n^{2/3}\sim 0.001$ keV
cm$^2$ (assuming $n\sim 1$ cm$^{-3}$) is 5 orders of
magnitude lower than the typical value of $\sim 100$ seen in cluster
cores.  Of course a hot medium may already be present but undetectable
in $z\sim 2$ QSO hosts, yet evidence for a significant warm phase
($T\sim 10^5$\,K), i
as might be expected at hot/cold interfaces, was not
uncovered in companion work \citep{QPQ3} and our stack does not show
statistically significant \ion{N}{5} or \ion{O}{6} absorption.
Our results suggest that a massive IGrM/ICM may not be in 
place at $z\sim 2$.
%% JFH The sentence below about quasar feedback is not correct. Models
%% of IGrM/ICM formation have argued that AGN feedback plays a role, but this is
%% usually believed to be a process involving radio jets, i.e. radio mode
%% feedback. There is some work on quasar mode kinetic feedback and its 
%% impact on galaxy formation, but most of that is in the quenching literature
%% I believe. 
% [CHAT]
%Regardless of its eventual fate, the cool gas must be
%eliminated from the IGrM/ICM.  
%%JFH I don't think you want to say removed, as that suggests ejected or expelled. 
%% The cold gas could just no longer exist, because it becomes hot or because
%% transport physics prevents cold clouds from surviving once most of the gas is hot
%% and/or the cold clouds don't form anymore. 
%% [CHAT -- Is eliminated ok?  Add text on the rest]
Recent models of IGrM/ICM formation have argued that
quasar feedback plays a critical role \citep{mccarthy10}, but
we note no influence of the quasar on gas on scales of
tens to several 100\,kpc.  We do find, however, that our (crude)
metallicity estimates for the CGM gas are consistent with the
enrichment level estimated for the IGrM/ICM \citep[e.g.][]{werner08}.
The processes that enrich the IGrM/ICM
may already be active at $z>2$.

%% May need to eliminate or at least significantly reduce the
%% following paragraph.
Previous work has suggested a causal connection between the $z\sim 2$
CGM and galactic-scale feedback \citep[e.g.][]{od08,steidel+10}.
Certainly, the presence of heavy elements distributed throughout
quasar hosts' halos demands an effective transport mechanism from the
sites of metal production.  Yet the properties of this CGM do not
immediately suggest an origin in violent outflows.  This optically
thick gas cannot have recently been subjected to significant heat
input, e.g.\ via shocks or conduction from an enveloping hot phase.
The large $W_{1334}$ values indicate motions on the order of a few
hundred km/s, but systems with the extreme kinematics required to
launch winds deep into the halo (widths of $\approx 1000\mkms$)
are relatively rare.  Presently, we favor scenarios where
the metals were formed primarily in lower mass satellites and then
ejected into the CGM by winds or dynamical stripping during infall
\citep[e.g.][]{shen+11}.  Ultimately, we will test these and other
scenarios with higher dispersion measurements of the gas metallicity
and kinematics.

\acknowledgements

The authors gratefully acknowledge the support which enabled these
observations at the Keck, Las Campanas, Calar Alto, and Apache Point
Observatories.  JXP is supported by NSF grant AST-1010004.
We thank N. Crighton, J. Tumlinson, and J. Werk for sharing 
results on the CGM prior to publication.
  Much of the data presented herein were
  obtained at the W.M. Keck Observatory, which is operated as a
  scientific partnership among the California Institute of Technology,
  the University of California and the National Aeronautics and Space
  Administration. The Observatory was made possible by the generous
  financial support of the W.M. Keck Foundation. The authors wish to
  recognize and acknowledge the very significant cultural role and
  reverence that the summit of Mauna Kea has always had within the
  indigenous Hawaiian community. We are most fortunate to have the
  opportunity to conduct observations from this mountain. 
  JXP acknowledges support from a Humboldt visitor fellowship to the Max
  Planck Institute for Astronomy where part of this work was
  performed. 

%{\bf [JFH: add SDSS acknowledgments.]}
%% JFH Well if we are crunched for words, maybe we have to omit them. 

%\bibliographystyle{/u/xavier/paper/Bibli/apj}
%\bibliography{/u/xavier/paper/Bibli/allrefs}

%\input{../Tables/tab_qpq5_summary.tex}
%\input{Tables/tab_qpq5_summary_sub.tex}

\begin{deluxetable*}{lccccccccccccccc}
%\rotate
\tablewidth{0pc}
\tablecaption{Sample Summary\label{tab:qpq5_summ}}
\tabletypesize{\scriptsize}
\tablehead{\colhead{F/G Quasar} &
\colhead{\zfg} &
\colhead{B/G Quasar} & 
\colhead{\zbg} &
\colhead{\rphys} &
\colhead{Spec.} & \colhead{\zlya$^a$} &
\colhead{log \nhi} &  
\colhead{$W_{1334}$} &
\colhead{$W_{1548}$} &
\colhead{flg$_{\rm OT}^b$} &
\\
&&& & (kpc) &&&& (\AA) & (\AA) &&}
\startdata
J002126.10$-02$5222.0&$2.692\pm 0.006$&J002123.80$-02$5210.9&3.291&299&BOSS&2.7010&$<18.80$&&$< 0.44$& 0&\\
J003423.06$-10$5002.0&$1.836\pm 0.003$&J003423.40$-10$4956.3&1.948& 67&LRISb&1.8350&$<18.70$&$< 0.09$&& 0&\\
J014216.40$+00$2328.5&$2.713\pm 0.006$&J014214.75$+00$2324.2&3.349&208&SDSS&2.7089&$<19.00$&&$< 0.15$& 0&\\
J022517.68$+00$4821.9&$2.727\pm 0.006$&J022519.50$+00$4823.7&2.818&226&GMOS&2.7302&$<18.80$&$< 0.11$&$0.44\pm0.12$& 0&\\
J074031.15$+22$4616.1&$2.334\pm 0.008$&J074029.77$+22$4557.2&2.647&228&BOSS&2.3297&$18.95\pm0.30$&&$<0.38$& 1&\\
J075009.25$+27$2405.2&$1.771\pm 0.003$&J075008.26$+27$2404.5&1.802&115&LRISb&1.7691&$<18.90$&$< 0.12$&& 1&\\
J075259.81$+40$1128.2&$1.883\pm 0.003$&J075259.13$+40$1118.2&2.121&110&LRISb&1.8830&$18.80\pm0.30$&$0.28\pm0.03$&$< 0.32$& 1&\\
J075435.39$+48$0631.6&$2.510\pm 0.003$&J075437.67$+48$0611.6&3.124&255&BOSS&2.5069&$<18.50$&&$< 0.46$& 0&\\
J080049.89$+35$4249.6&$1.981\pm 0.003$&J080048.73$+35$4231.3&2.066&201&LRISb&1.9824&$18.70\pm0.40$&$0.65\pm0.02$&& 1&\\
J081420.37$+32$5016.1&$2.173\pm 0.008$&J081419.58$+32$5018.6&2.213& 88&GMOS&2.1790&$18.80\pm0.30$&$0.31\pm0.04$&$< 0.16$& 1&\\
\enddata
\tablenotetext{a}{Redshift characterizing the \lya\ absorption in the spectrum of the background quasar.}
\tablenotetext{b}{Assessment of whether the system is optically thick at the Lyman limit ($-1$=Thin; 0=Ambiguous; 1=Thick).}
\tablecomments{[The complete version of this table is in the electronic edition of the Journal.  The printed edition contains only a sample.]}
\end{deluxetable*}

%%%%%%%%%%%%%%%%%%%%%%%%%%%%%%%%%%%%%%%%%%%%%%%%%%%%%%%%%
%%%%%%%%%%%%%%%%%%%%%%%%%%%%%%%%%%%%%%%%%%%%%%%%%%%%%%%%%
%%%%%%%%%%%%%%%%%%%%%%%%%%%%%%%%%%%%%%%%%%%%%%%%%%%%%%%%%

\end{document}